# Cognitive Convergence: Deep Similarities Between Large Language Models and Human Cognition

Chandra Sripada[1,2,3], Richard Lewis[3,4,5]

1. Department of Philosophy, University of Michigan
2. Department of Psychiatry, University of Michigan
3. Weinberg Institute for Cognitive Science, University of Michigan
4. Department of Psychology, University of Michigan
5. Department of Linguistics, University of Michigan

## Abstract

LLMs are widely regarded as alien intelligences, systems whose cognitive operations are fundamentally unlike our own. Apparent similarities to human cognition are therefore often seen as the result of anthropomorphic projection. We argue that this framing is mistaken. LLMs clearly differ from humans in important respects, including their physical substrate, learning history, and the environments with which they interact. These differences make it all the more striking that contemporary LLM-based systems converge with human cognition on a number of principles of cognitive organization with longstanding support in cognitive science. We identify structural correspondences across five dimensions: inferential organization, computational architecture, representational structure, prediction-driven learning, and reinforcement-learning-like mechanisms supporting goal-directed action. These correspondences support a broader model of intelligent cognition in which core principles long used to explain human intelligence also characterize contemporary LLM-based systems.

*“Traditionally, the term ‘AI’ has been used as an acronym for artificial intelligence. But it is perhaps better to think of it as an acronym for alien intelligence.” - Yuval Noah Harari*[1]

# 1. Introduction: From Alien Intelligence to Cognitive Convergence

Theorists disagree sharply about the capabilities of large language models (LLMs). Some dismiss them as “stochastic parrots”[2] while others argue they have already achieved artificial general intelligence[3,4] or likely will in the near future[5]. Yet many on both sides converge on a deeper assumption: whatever LLMs are, their cognition is fundamentally alien—so unlike our own that apparent similarities mostly reflect anthropomorphic projection.[6–11] This assumption is not universal, but it has become prominent across popular media, technology commentary, public intellectual discourse, and parts of the behavioral and brain sciences.

We argue that this framing is mistaken—or, at the very least, greatly overstated. LLMs plainly diverge from humans in critical respects, including physical substrate, learning history, and the environments with which they interact. The mistake is to treat these contrasts as decisive evidence of fundamentally different *cognitive organization*. Our central claim is that, despite salient differences, humans and contemporary LLM-based systems converge on several core principles of intelligent cognition with longstanding support in cognitive science.

We begin by reviewing systematic parallels at the level of outputs between LLMs and humans, while arguing that behavioral resemblance alone cannot settle the issue. We then examine five linked dimensions of cognitive organization. Each addresses a general problem faced by intelligent systems and has independent motivation within cognitive science: inferential organization; computational architecture; representational structure; prediction-driven learning; and reinforcement-learning-like mechanisms for goal-directed action. We identify structural correspondences across all five dimensions that, when taken together, substantially weaken the alien intelligence framing.

The resulting picture is one of humans and LLM-based systems as “cognitive cousins”—importantly different systems that nevertheless instantiate some of the same basic principles of intelligent cognition. Their convergence despite extensive differences in substrate and learning history is especially striking, suggesting that principles long used to explain human intelligence apply more generally across the biological–artificial divide.

# 2. Behavioral Parallels and the Need for Deeper Mechanistic Comparison

The case for viewing LLMs as alien intelligences often begins with conspicuous behavioral anomalies—for example, some models cannot reliably count the r’s in “strawberry”[12] or answer “Who is Marco’s wife?” after learning during training that Lina is married to Marco[13]. But in addition to noting differences, we should also ask whether equally systematic *similarities* appear at the behavioral level. They do. The important evidence is not merely that LLMs sometimes attain human-level accuracy, but that their performance often exhibits a nuanced human-like profile: the same items tend to be easy or difficult, the same response alternatives are preferred, and the same experimental manipulations increase the likelihood of error. The picture that is emerging is one in which the behavioral profiles of LLMs and humans align in nontrivial ways.

Consider examples from language processing. A singly center-embedded sentence such as “The teacher the student saw is happy” is manageable. But adding another layer—“The teacher the student the driver saw hit is happy”—quickly becomes difficult, because the comprehender must keep several unfinished dependencies active at once.[14] “The cotton clothing is made of grows in Mississippi” is challenging to parse but “The cotton *that* clothing is made of grows in Mississippi” is easy; the former invites an incorrect initial parse (a “garden path”) that must be revised. These constructions[15] as well as others involving similarity-based interference[16] (in which a word with features similar to a target serves as a distractor) and depth shifts[17] (in which pragmatic expectations override syntactic analysis) pose well-established challenges to humans. Language models likewise show a similar pattern of systematic sensitivity to these constructions.[18–21]

Or consider serial memory. When humans study a list of items and later recall them, they do not retrieve each position equally well: the first and last items are favored over the middle, and recalling one item tends to cue nearby items, especially those that followed it in the original sequence.[22,23] LLMs show closely related retrieval biases. Relevant information is often easier for them to use when placed near the beginning or end of a context than when placed in the middle.[24] Additionally, when presented with stimuli that match parts of previously presented sets, their outputs, as well as internal attention patterns, exhibit temporal contiguity and forward asymmetry effects resembling those found in human episodic recall.[25]

Similar correspondences are observed in other domains, including analogical reasoning, theory of mind, semantic cognition, cognitive tasks, and decision-making.[26–33] Across all these domains, LLMs and humans show similar relative item difficulty, option preferences, and error patterns.

A particularly vivid example comes from visual search. In humans, disjunctive search is relatively easy: a green circle among red circles “pops out,” because a single feature uniquely identifies the target. Conjunctive search is harder: a green L among red Ls and green Ts requires binding color to shape, because neither “green” nor “L” alone picks out the target.[34] Recent work on vision-language models, which are trained on visual inputs in addition to human language, finds the same qualitative asymmetry.[35] The same pattern appears in rapid numerosity judgments. Humans can subitize small sets, but performance deteriorates beyond roughly four to six items unless they can count serially.[36] Multimodal and text-to-image models show similar performance profiles: they perform well for small numerosities but their performance rapidly deteriorates for larger sets.[35]

One could argue that these similarities are not interesting—they emerge simply because LLMs have been trained to mimic patterns found in human text. But note that this explanation already fails to provide a straightforward account of the serial memory effects and visual search effects noted above, which the researchers who identified these effects attribute to shared underlying mechanisms, e.g., induction head-like mechanisms in the temporal contiguity and forward asymmetry effects[25] and feature binding mechanisms in conjunctive search[35]. The more important point, perhaps, is that there is limited utility in simply trying to count and tabulate behavioral differences and similarities. If we focus exclusively on behavior, we risk missing theoretically important commonalities or differences at the level of architecture and mechanism. Thus, in the following sections, we seek to take a more principled approach to the issue: We identify a number of deeper and more theoretically important dimensions in which LLM processing principles bear striking resemblances to human cognition.

# 3. Inferential Organization: Dual-Process Structure

Inferential organization poses a general problem for intelligent systems: how to combine fast, experience-shaped responses with slower, flexible reasoning about novel situations.[37,38] A longstanding and influential answer in cognitive science is dual-process organization: Type 1 processes are fast, automatic, and shaped by prolonged prior experience, whereas Type 2 processes are slower, sequential, working-memory-dependent, and capable of rapidly interpreting novel task structures.[39–41] Perhaps surprisingly, transformer systems, pretrained through sequence prediction and sometimes later refined through reinforcement learning (as discussed below), develop an analogous division of cognitive labor.

Two related but distinct mechanisms underwrite this resemblance. The first is the contrast between in-weight knowledge and in-context inference.[42–44] Pretraining gradually installs regularities into the model's parameters ranging from factual associations ("The capital of France is Paris") to commonsense regularities ("If you drop an object it will fall").[45–47] This is the LLM analogue of slow, experience-dependent learning that compiles information into model weights. But transformers also acquire the ability to infer a temporary task structure from the current context, without any update to those weights. Given examples such as "aaab → baaa" and "ccct → tccc," a model can infer the abstract transformation and apply it to a new case such as "vvvs → svvv." In such cases the model is not simply retrieving a stored fact. It is using the prompt as a temporary workspace in which a new rule can be inferred and applied online.[43,48,49]

The second mechanism is the contrast between direct processing and chain-of-thought-mediated reasoning.[50–53] In direct processing, the model maps the prompt relatively immediately onto an answer. This often works when the problem exhibits familiar patterns, but it can fail in cases that require inhibiting an intuitive response. Mitchell (2025) illustrates the point with a puzzle: "Julia has two sisters and one brother. How many sisters does her brother Martin have?" A direct response may be drawn toward "two" by the locally salient phrase "Julia has two sisters." In chain-of-thought-mediated reasoning, by contrast, the model first generates intermediate tokens that become part of the context for subsequent prediction: "Julia has two sisters, so there are three girls in the family: Julia and her two sisters. Martin is the brother, so Martin has three sisters." The model uses the ability to generate linguistic tokens to create an extended cognitive scratchpad, converting a mechanism that performs one-pass pattern completion into a mechanism for serial inference.

Importantly, both mechanisms underpinning dual-process structure in LLMs are also strongly psychologically plausible. Russin and colleagues showed that the in-context/in-weight distinction explains many fundamental effects observed in the cognitive science of learning, such as compositional generalization patterns, blocking-versus-interleaving curriculum effects, and the flexibility-retention trade-off.[44] Other researchers have linked the distinction to conflict tasks, such as the Stroop and flanker tasks, in which an in-context task instruction competes with an in-weight automatic response tendency.[54] De Varda and colleagues linked chain-of-thought-mediated reasoning in LLMs with human reaction times. They studied seven categories of tasks encompassing roughly 150 problems and demonstrated that the number of internal thinking tokens used by LLMs to solve a problem is strongly predictive of thinking times for humans in those same tasks.[55]

Dual-process structure is widely recognized as a central organizational feature of human cognition.[39–41] It is thus remarkable that transformer LLMs recognizably exhibit a parallel dual-process distinction, underpinned by a pair of mechanisms with strong psychological plausibility. LLMs also provide cognitive science with a new model for explaining the *emergence* of dual processes in the human mind, grounded in learning mechanisms with independent psychological motivation (i.e., prediction and reinforcement learning, which are discussed below).

# 4. Computational Architecture: Production-System-Like Organization

Flexible intelligence requires a computational organization that can keep track of the current situation and selectively apply and compose operations in response to its relevant features. In cognitive science, production-system architectures have long offered an influential account of how human cognition meets this challenge.[56–59] Transformer LLMs exhibit a strikingly similar organization: in both systems, a persistent representational workspace is iteratively modified by a large population of context-sensitive operators.

To see what is distinctive about this organization, it helps to contrast it with the serial instruction-following approach familiar from ordinary digital computers. These systems operate with a "fetch-execute" cycle: a program counter identifies the next instruction, the processor fetches it, decodes it, executes it, and then advances to the next instruction.

Production systems represent a very different, more flexible style of computation: they operate with a massively parallel “recognize-act” cycle.[60] A production system maintains a working memory, a structured representation of the current situation and immediate goals, and a long-term memory consisting of a large set of condition-action rules, or “productions.” Each production has the form: *if* some condition is present in the current working memory state, *then* perform some one or more actions. An action typically modifies the contents of working memory, retrieves information, updates a variable, and so forth. Control does not follow fixed paths laid out by a program; it instead emerges from the interaction between a shared workspace and a population of rules that activate whenever their conditions are satisfied. Furthermore, the match, and in some theories also the execution, of productions is a parallel process.[58]

The core ideas behind production systems emerged in the early work of Emil Post, who abstractly characterized computation in terms of the repeated application of conditional transformation rules.[61] In cognitive science, production systems gained prominence through the work of Allen Newell, Herbert Simon, and John Anderson, among others.[57,62–64] Newell in particular argued for “unified theories of cognition”: general cognitive architectures capable of explaining perception, memory, reasoning, learning, and action in a shared computational framework.[57] Production systems were attractive for this purpose because they readily account for three features that appear to be necessary for sophisticated cognition: modularity, recombination, and generality.[56]

A production is a small piece of procedural knowledge: a context-sensitive transformation that says what to do when certain conditions obtain. Such units are modular because each rule makes only a local, targeted change, and individual rules can be learned and revised without rewriting the whole system. They are composable because complex skills can be built by combining many such rules in sequence or coordinating them in parallel. And they are general because productions are a highly flexible medium for computation. At a formal level, systems of condition-action rules, operating over a sufficiently rich working memory, are computationally universal in the Church-Turing sense: they can implement any well-specified procedure.[61,65]

Over the last several decades, production-system architectures have been used in cognitive science to model a wide range of human cognitive phenomena, including problem solving and reasoning[57,62,66], skill acquisition[67], memory retrieval[68,69], multitasking[70], language comprehension[14], and perceptual-motor control[71]. They are among the few approaches that attempt to cover a very wide range of capacities and empirical phenomena within a single framework.[57,59] They stand out for their explanatory breadth, but also for experimental precision. ACT-R and related architectures generate fine-grained predictions about reaction times, accuracy, learning curves, error patterns, and, in some cases, brain activation patterns.[72–74] It is therefore fair to say that production-system architectures are among the preeminent frameworks in cognitive science for detailed computational modeling of how cognition unfolds over time, and there are few viable alternatives.[75,76]

LLMs are sometimes portrayed as ‘black boxes,’ but their broad computational architecture is well understood[77–79], and at that level they bear a striking structural resemblance to production-system architectures—a correspondence that remains underappreciated by theorists. A central architectural component in transformers is the *residual stream*, a high-dimensional vector associated with each token position that persists through the layers of the network.[78,79] Attention blocks and MLP blocks read from this stream and write back to it, each block making an incremental update rather than replacing the entire representation.[78–82]

This residual strategy was popularized by He and colleagues in their seminal work on ResNets[83], where they addressed the “degradation problem”: as standard neural networks became deeper, adding more layers often yielded worse performance. Their solution was to add identity skip connections, so that each block learned an incremental update relative to the representation it received rather than learning a complete rewrite. The result is an architecture biased toward modular, incremental modifications that compose across layers. This pattern closely mirrors a key feature that made production systems attractive in cognitive science: individual productions make local, targeted changes to a persistent memory state, and these changes compose to yield useful computations.[56,57,59,72]

The production-like nature of transformers is also visible by examining the specific computational operations performed at their two key components, attention heads and MLP units. Attention heads implement a graded, differentiable version of condition matching and information routing. A head computes queries, keys, and values from the residual stream. For each token position, query-key comparisons determine how strongly that position attends to other positions. The resulting attention weights determine a weighted combination of value vectors that is combined and ultimately written back into the residual stream. Thus, an attention head has a production-like structure: when it detects a certain relational pattern in the current context, it uses that match to construct an update to the residual stream.

The MLP blocks in transformers also have a production-like interpretation, often characterized explicitly in similar key-value terms.[80–82] An MLP reads the residual stream at a token position, uses its first linear layer and nonlinearity to compute degrees of match to learned feature directions, or “keys,” and then uses those activations to form a weighted sum of learned output vectors, or “values,” which constitutes the update to the residual stream. MLPs are thus naturally understood as implementing graded condition-action operations: if the current representation contains such-and-such features, then add such-and-such vectorial modifications to the workspace.

Recent computational work further demonstrates intimate connections between production systems and transformer architectures. Smolensky et al. (2025) created a production rule language that enables construction of arbitrary systems of discrete production rules.[84] They also built a corresponding compiler that translates the rules written in this language into the parameters of a transformer and showed that the transformer successfully implements the specified productions (they additionally proved the Turing-completeness of the language).

To be clear, the claim is not that transformer-based LLMs literally implement classical production systems. Classical productions are typically discrete, symbolic, and largely hand-coded; transformer operations are dense, graded, and learned. The relevant similarity is instead at a higher level of computational organization: in both systems, intelligence depends on a persistent shared workspace that is modified by a large population of conditional operators whose effects can be flexibly composed to yield useful computations, a strategy that has remarkable generality across task domains.

Some may accept the deep similarities between transformer-based LLMs and production systems in cognitive science but resist describing this as a case of convergence, since the transformer architecture is deliberately designed rather than emergent through learning. The relevant design choices, however—including ResNet-style residual connections[83] and the key-query structure of attention[77]—were not introduced to mimic production system architectures; they were independently motivated by the aim of improving model performance. The convergence lies in the fact that a series of largely independent choices in AI and computer science produced an LLM architecture that bears important similarities to production systems, one of the most influential approaches to modeling human cognition.

# 5. Representational Structure: Alignment, Contravariance, and Canalization

Shared computational organization leaves open whether humans and LLMs also encode and transform task-relevant information in similar ways. A growing body of evidence suggests that they do, revealing important correspondences between the representations and computational procedures operative in LLMs and those found in the human mind and brain. These correspondences also raise an explanatory question: why should such different systems converge on related internal solutions?

One route to identifying such representational correspondences is mechanistic interpretability, which seeks to characterize a trained network's internal computations in terms of human-interpretable features, circuits, and algorithms, and to test whether those proposed mechanisms causally contribute to the model's behavior through interventions[85–88]. Recent work using these methods has identified causally relevant internal states within LLMs that correspond to a range of constructs from theoretical linguistics, including features that encode grammatical number[89,90], relative-clause boundaries[91,92], bindings between entities and their attributes[93], and abstract structure shared across filler–gap dependencies[94]. Indeed, Kryvosheieva and colleagues identified internal units in LLMs implicated in 67 syntactic phenomena.[95] Additionally, the internal patterns of attention in LLMs show signatures of linguistic *similarity-based interference*, which is a central phenomenon in computational theories of human sentence processing.[14] Metrics that quantify the degree of interference in LLMs are highly predictive of human word-by-word eye fixation durations and reading times[96]—even though these models were, of course, not trained with the objective of fitting eye fixation or human reading time data. These findings indicate that LLMs do not merely produce fluent, human-like outputs; their internal states also approximate key constructs from theories of human language processing (for a detailed discussion of this point, including the relevant sense of "approximate," see[97]).

A complementary approach uses representational similarity analysis, neural encoding models, and related alignment methods to compare the geometry and predictive content of internal activation patterns across artificial and biological systems[98,99]. Using these approaches, Schrimpf et al. (2021) found that, across 43 language models and three neural datasets, the models most successful at next-word prediction also best predicted human neural responses and behavioral measures of comprehension difficulty, in some datasets approaching the estimated noise ceiling[100] (for related convergent results, see[101–103]).

A particularly notable form of model-brain alignment concerns not merely the content of representations but their ordering across successive processing stages. Mischler et al. (2024) compared layerwise activations from LLMs with intracranial recordings from eight neurosurgical patients as they listened to natural speech.[104] Model depth mapped systematically onto a cortical gradient indexed by distance from primary auditory cortex: relatively shallow layers best predicted activity at nearby sites, while progressively deeper layers best predicted sites farther along the auditory and language-processing pathway. This pattern resembles hierarchical correspondences previously demonstrated in vision and audition, where task-optimized deep networks predict responses across successive stages of the ventral visual pathway[105] and auditory cortical processing pathway[106], progressing from relatively local, stimulus-bound features at earlier stages toward more integrated, invariant, and task-relevant representations at later stages.

One influential theoretical explanation for these correspondences at the level of representations and procedures appeals to *contravariance*: as tasks become more demanding, the space of viable solutions narrows[107]. That is, for relatively easy tasks, many internal solutions may perform adequately; as tasks become more demanding, the range of viable solutions narrows.. If human-level cognitive skills represent one of the hardest optimization targets, then systems that display similar skills—despite differences in training data or substrate—will tend to converge on similar internal solutions. On this view, the observed representational alignment between LLMs and humans is not an accident, but a consequence of limitations in the space of possible solutions.

We propose a second, complementary explanation. Transformers are not just any kind of neural network. As was discussed in §4, they are networks that exhibit a highly distinctive production system-like architecture centered around a shared residual stream that is sequentially modified across layers by a large population of conditional rules, a mode of computational organization that is plausibly shared by the human mind/brain. This suggests an additional explanation that invokes *architectural canalization*.

The core idea is that two systems built around similar production-system-like architectures and then trained to solve similarly demanding cognitive tasks will tend to converge on similar solutions, even if exposed to ostensibly different kinds of training regimes.[108,109] This is because the presence of a production-system-like architecture strongly biases the kinds of operations that tend to emerge within it. Such architectures favor operations whose outputs make incremental changes to a persistent shared workspace. This, in turn, promotes reusable and composable rules that can be recombined across tasks and domains rather than procedures that repeatedly rewrite the entire representational state.

Contravariance and architectural canalization are best seen as complementary explanations for LLM/human representational alignment and other processing-level commonalities. Contravariance emphasizes that difficult problems admit a comparatively narrow range of viable solutions. Architectural canalization adds a further constraint: within that range, shared architectural biases make some specific solutions more likely to emerge than others.

# 6. Prediction-Driven Learning

A general problem for intelligent systems is how to extract structure from experience and revise internal models as new evidence arrives. Prediction provides a powerful solution: systems generate expectations and use mismatches between expectation and observation to update their representations. In cognitive science, this principle has longstanding support across perceptual judgment, motor control, language, and higher-level cognition[110–114], leading to proposals that it represents a fundamental principle of human cognition[115–117]. Decades of work in control theory provide the statistical motivation: when tracking hidden states via noisy sensory signals, prediction can substantially improve estimation by combining an internal model of how the world evolves with incoming sensory feedback (e.g.,[118]).

Autoregressive LLMs are explicitly trained through a prediction-and-correction loop: given a preceding context, they make a probabilistic prediction of the next token, compare the prediction with the observed continuation, and use gradient descent to adjust their parameters so as to reduce future prediction error. Although prediction-based learning is therefore built into LLMs, this nevertheless reflects a case of convergence when we consider the sequence of choices that led to its adoption. Across decades of research, AI and computer-science researchers developed and combined statistical prediction, error-driven learning, language modeling, and large-scale training because these methods improved model performance[119–121], not mainly because they sought to mimic theories of human cognition. Thus, an independently pursued engineering trajectory ultimately made prediction-based learning central to LLMs, converging on a principle that cognitive scientists have identified as fundamental to the operation of the human mind.

This convergence on prediction becomes more consequential when combined with the shared production-system-like architecture described above. When systems with partly overlapping computational architectures are shaped by analogous predictive pressures, architectural canalization can push them toward similar representational structures and conditional operations. This may help explain the correspondences between humans and LLMs in behavioral outputs, inferential organization, and representational landscapes reviewed in the preceding sections.

This commonality nevertheless coexists with an enormous difference in *sample efficiency*. LLMs are trained on internet-scale data that dwarf the linguistic experience available to a human child. GPT-3, for example, was trained on 300 billion tokens. By contrast, the developmental estimates used in the BabyLM project place a child's cumulative linguistic exposure by age twelve at approximately 24–84 million words[122,123] (which for the current purposes can be taken to approximate tokens), a gap of roughly three to four orders of magnitude.

This gap establishes a clear difference in the learning process, but it does not by itself establish a corresponding difference in the learning target, i.e., the set of representations and computational procedures acquired. Drawing on Locke's famous question of how the mind comes to be "furnished," we should be careful to distinguish the process of furnishing the mind from the arrangement that its tables, desks, and chairs ultimately assume.[124] Current LLMs are clearly "furnished" much less efficiently than human children are: they require much more linguistic experience to acquire robust competence. But it would be a mistake to infer from this acknowledged fact that the functional and representational organization they eventually assume must therefore be radically different.

The distinction is illustrated by studies asking whether convolutional structure can be learned rather than built into an architecture. Ingrosso and Goldt (2022) trained an initially fully connected network—one possessing neither local connectivity nor tied weights—on a translation-invariant visual discrimination task.[125] Given sufficiently extensive training, its hidden units developed localized receptive fields that tiled the input space in a way that strongly resembled the filters learned by a convolutional network trained on the same task. Despite this convergence in terms of eventual learning target, the respective networks still strongly differ in terms of sample efficiency, as learning-curve analyses demonstrate that fully connected networks that lack built-in convolutional structure require dramatically more training data.[126] The evidence reviewed in the preceding sections suggests an analogous phenomenon in LLMs. Their learning process is dramatically less sample efficient, but the organization they eventually acquire resembles important aspects of the human mind-brain: a sophisticated system of representations and conditional operations supporting dual-process inferential organization, linguistic fluency, reasoning, and other features of complex human cognition.

It is a further empirical question whether the low sample efficiency of present-day LLMs is a deep and unavoidable feature or instead reflects the absence of a relatively small number of critical learning mechanisms. Several lines of research support the latter possibility. Stronger structural inductive biases, for example, convolutional locality assumptions in early visual processing, can substantially improve data efficiency.[126] Learned internal world models allow "offline" training on imagined trajectories, dramatically reducing training time.[127] Curiosity-driven agents can actively seek observations for which their predictions are poor, rather than passively consuming randomly supplied data.[128] Multimodal grounding may also make limited linguistic input more informative. Vong et al. (2024), for example, trained a neural model on just 61 hours of paired visual and linguistic experience recorded from a single child.[129] On an in-distribution word-referent classification task, the model achieved 61.6% accuracy, approaching the 66.7% accuracy of CLIP, a web-scale vision–language model trained on 400 million image-text pairs.

Overall, we can think of humans and LLMs as climbing the same mountain, both relying on prediction-based equipment. Humans follow a straighter, more direct route, aided by mechanisms such as evolved inductive biases and multimodal feedback. Present-day LLMs follow a longer, more circuitous, and more wasteful route through vast corpuses of text. But circuitousness alone is not evidence of an alien destination, particularly when the two routes appear to converge on similar representations and modes of inferential organization.

# 7. Reinforcement Learning and Reinforcement Learning and Agentive Control

Prediction alone does not fully explain goal-directed control. Prediction learning can equip a system with sophisticated representational structures, but an account of agency must additionally explain how candidate actions and behavioral trajectories are evaluated, selected, maintained, and revised in light of their outcomes. Reinforcement learning (RL), understood broadly as a computational approach in which evaluative feedback about action outcomes shapes future action selection, provides a general framework for explaining these functions.[130]

In computational cognitive science, RL has become an influential unifying framework for understanding human decision-making and action selection[131–134]. Importantly, however, this does not imply that a single RL mechanism confers agency on the system. Instead, human reward-guided action appears to recruit multiple partly dissociable processes of valuation, control, and metacontrol. For example, actor-critic models have been used to model action selection in basal-ganglia circuits[135], and model-free temporal-difference accounts have been used to model habit learning and habitual control[136]. RL and value-based models have also been used to characterize executive processes, such as how cognitive control is allocated[137], how planning strategies are selected[138], and how people choose among reasoning heuristics[139]. Still other RL-related processes are proposed to contribute to homeostatic motivations such as hunger and thirst[140,141] and affective phenomena such as emotions[142] and moods[143]. Importantly, where distinct action controllers favor incompatible actions, for example, conflict between model-based and habitual RL controllers[144,145], their relative influence must be coordinated. Arbitration among these systems has itself been modeled in value-based and RL-like terms.[146,147]

These observations illustrate a key point: With human action selection and decision-making, RL constitutes an *infrastructure* of interacting routines. Its importance for agency does not lie in the existence of a single RL mechanism that converts the system into an agent. It lies in the way a general reward-guided learning schema can be instantiated across many distinct control problems. At each locus, the relevant variables change: what actions are available, what counts as feedback, the time horizon which must be tracked, how reward is represented, and what forms of failure are encountered. The unity is therefore at a formal level in which RL provides an abstract computational grammar for agency. This grammar includes theoretically well-defined targets for learning—among them value functions, policies, and environment models—and their formal relationships.[130] There are, however, importantly different configurations constructed from this grammar distributed across different sites of control within the system.

In contemporary LLM-based systems, RL is a central, though not exclusive, framework for shaping goal-directed behavior, marking an important point of convergence with humans. But there is a key difference: the RL landscape is much more limited. Early approaches used reinforcement learning with human feedback (RLHF), a single turn form of RL, to bend model response policies toward outputs preferred by human evaluators.[148] More recent reasoning systems optimize the same general policy over longer reasoning trajectories, often using automatically verifiable outcome rewards and policy-gradient methods such as proximal policy optimization and related routines[52,149,150]. Policy-gradient methods have also been used to train LLM-based agents with access to additional environment-facing actions, such as searching the web, inspecting files, managing an external memory, delegating tasks to subagents, and so forth.[151,152]

While the achievements of RL in LLM-based systems are significant, for example, RL-trained systems reached gold-medal-level performance on the 2025 International Mathematical Olympiad[153,154], what is just as salient is what is currently missing. We do not observe a complex, hierarchical organization with partly dissociable processes of valuation, control, and metacontrol. We also see limitations in the duration and complexity of tasks that these systems can complete autonomously.[155] More seriously from the perspective of AI safety, RL-trained reasoning agents can engage in reward hacking or pursue proxy objectives rather than the intended task.[156,157] Also notable is the absence of analogues of homeostatic and affective RL mechanisms that shift the valuational landscape globally in response to internal state, situational cues, or signals about general reward availability. And yet, the key point is that none of this is strong evidence that artificial agency is alien in any fundamental computational sense. It shows instead that the RL infrastructure current deployed in artificial systems is incomplete, brittle, and underdeveloped.

Some of these limitations may be reduced through broader and more diverse training experience. Such experience may come from interaction with real-world environments[158] or from behavioral trajectories simulated using learned world models[127] or through self-play[159]. Another complementary avenue involves engineering tweaks and optimizations, for example building better scaffolds for memory, tool use, and long-horizon control[160–162]. Whether sufficiently broad experience and optimization pressure can induce a more complex RL infrastructure with functionally distinct controllers and metacontrol within a single model[163,164], or whether these capacities require deliberate architectural modifications, remains a critical open empirical question.

Other limitations may prove more stubborn. For example, during complex task performance, humans appear to receive extensive, dense, highly calibrated reward signals delivered by internal valuational networks that have been shaped over evolutionary and developmental time.[165] Artificial systems, in contrast, often receive impoverished, sparse rewards that are defined most easily in verifiable domains like game playing, math, and coding.[150,166] It may prove challenging to artificially engineer rich, highly calibrated reward signals in non-verifiable domains such as social cognition, scientific discovery, and artistic creation. Such domains—and more generally the problem of reward discovery and design—are areas of active research in the RL community.[167–169]

The key point, however, is that nascency and underdevelopment are not the same thing as alienness. The convergence lies at the level of a shared RL computational framework: both human and artificial agents use reward signals to learn policies that map environmental states to actions under constraints of limited knowledge, time, and resources. The RL infrastructure currently deployed in artificial systems tend to be limited and underdeveloped, but this reflects the current state of scaling and engineering—how frontier models are presently constructed, trained, scaffolded, and rewarded—not evidence of a wholly foreign kind of agency. If agency in humans and machines is realized, at least in part, through a shared family of RL-like computational processes, then RL represents an important point of continuity between biological and artificial intelligence. Here again, the alien intelligence framing loses significant traction under closer scrutiny.

# 8. Conclusion

The evidence reviewed here supports a model of cognitive convergence: despite differing from humans in important respects, contemporary LLM-based systems instantiate multiple core principles long used in cognitive science to explain human intelligence. Both exhibit production-system-like structure, in which a persistent representational workspace is repeatedly read and updated by a population of conditional operators, yielding flexible and composable computation. Both display dual-process organization, which enables both fast, compiled responses and slower, resource-intensive serial reasoning. Prediction-driven learning plays a central role for both in acquiring representational and procedural competence, while reinforcement-learning-like mechanisms shape and deploy this competence in the service of goal-directed control. Constraints imposed by demanding tasks (contravariance) and by learning that unfolds within a shared computational architecture (canalization), in turn, help explain convergence in internal representations and behavioral profiles. Some of these convergences (dual-process organization and representational structure) emerge through pretraining, whereas others arise from deliberately engineered features (transformer architecture, prediction-based learning, and reinforcement-learning-like mechanisms for goal-directed control), where convergence reflects engineering trajectories with mostly technical aims that arrived at principles independently identified as central to human cognition.

Humans and LLM-based systems clearly differ in critical respects, including in embodiment, input modalities, inductive biases, learning histories, and reward structures. But these extensive differences make their convergence across core principles of intelligent cognition all the more striking and theoretically informative. Overall, the correspondences reviewed here support the view that humans and LLM-based systems are importantly different members of a shared computational family. These systems are, in this sense, our cognitive cousins: different in substrate and learning history, yet recognizably related in core principles of cognitive organization.